\documentclass[twocolumn,prl,aps,superbib,tightenlines,floatfix]{revtex4}
\usepackage{amsfonts}
\usepackage{amsmath}
\usepackage{amssymb}
\usepackage{graphicx}
\begin{document}
\bibliographystyle{apsrev}

\title{Dynamical corrections to the DFT-LDA electron conductance in
nanoscale systems}
\author{Na Sai}
\affiliation{Department of Physics, University of California, San
Diego, La Jolla, CA 92093-0319}
\author{Michael Zwolak}
\affiliation{Physics Department, California Institute of
Technology, Pasadena, CA 91125}
\author{Giovanni Vignale}
\affiliation{Department of Physics and Astronomy, University of
Missouri, Columbia, MU 65211}
\author{Massimiliano Di Ventra\cite{MD}}
\affiliation{Department of Physics, University of California, San
Diego, La Jolla, CA 92093-0319 }
\date{\today}

\begin{abstract}

Using time-dependent current-density functional theory, we derive
analytically the dynamical exchange-correlation correction
to the DC conductance of nanoscale junctions. The
correction pertains to the conductance calculated in the
zero-frequency limit of time-dependent density-functional theory
within the adiabatic local-density approximation.  In particular, we
show that in linear response the correction depends non-linearly on
the gradient of the electron density; thus, it is more pronounced for
molecular junctions than for quantum point contacts. We provide
specific numerical examples to illustrate these findings.
\end{abstract}
\maketitle

Recent attempts to apply static density-functional theory~\cite{HK_KS}
(DFT) to electronic transport phenomena in nanoscale conductors have
met with some success.  Typical examples are atomic-scale point
contacts, where the conductance, calculated with DFT within the
local-density approximation (LDA), is found to be in excellent
agreement with the experimental values~\cite{Jan}.  When applied to
molecular junctions, however, similar calculations have not been so
successful, yielding theoretical conductances typically larger than
their experimental counterparts~\cite{DiVentra00}. These discrepancies
have spurred research that has led to several suggested explanations.
One should keep in mind, however, that these quantitative comparisons
often neglect some important aspects of the problem.  For instance,
the experimentally reported values of single-molecule conductance seem
to be influenced by the choice of the specific experimental
set-up~\cite{Reed,Lindsay,Reichert,Tao}: different fabrication methods
lead to different conductances even for the same type of molecule,
with recent data appearing to be significantly closer to the
theoretical predictions~\cite{Tao} than data reported in earlier
measurements~\cite{Lindsay2}.  In addition, several current-induced
effects such as forces on ions~\cite{DiVentra04} and local
heating~\cite{DiVentra03} are generally neglected in theoretical
calculations. These effects may actually generate substantial
structural instabilities leading to atomic geometries different than
those assumed theoretically. However, irrespective of these issues,
one is naturally led to ask the more general question of whether
static DFT, within the known approximations for the
exchange-correlation (xc) functional, neglects {\it fundamental}
physical information that pertains to a truly non-equilibrium
problem. In other words, how accurate is a static DFT calculation of
the conductance within the commonly used approximation for the xc
functional, LDA, compared to a time-dependent many-body calculation in
the zero-frequency limit?

In this Letter we provide an answer to this question. Specifically, we
seek to analytically determine the correction to the conductance
calculated within the static DFT-LDA approach and illustrate the
results with specific examples.  A few recent attempts were made in
this direction.  For instance, Delaney {\it et al.}~\cite{Delaney}
used a configuration-interaction based approach to calculate currents
from many-body wavefunctions. While this scheme seems to yield a
conductance for a specific molecule of the same order of magnitude as
found in earlier experiments on the same system, it relies on strong
assumptions about the electronic distribution of the particle
reservoirs~\cite{Delaney}. Following Gonze {\it et al.}~\cite{Gonze},
Evers {\it et al.}~\cite{Ever} suggested that approximating the xc
potential of the true nonequilibrium problem with its static
expression is the main source of discrepancy between the experimental
results and theoretical values. However, these authors do not provide
analytical expressions to quantify their conclusion.

\begin{figure}
\includegraphics*[width=5.5cm]{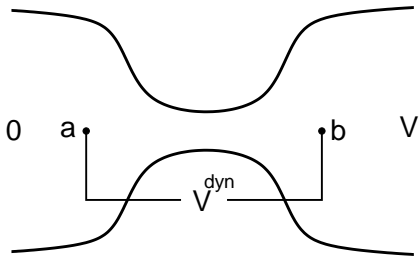}
\caption{Schematic of a nanoscale junction between electrodes with
applied bias $V$. Due to dynamical exchange-correlation
(xc) effects there is an xc field which gives rise to an additional
voltage drop $V^{\rm dyn}$ compared to the electrostatic potential
difference.}
\label{model}
\end{figure}

Our system is the nanojunction illustrated in Fig.~\ref{model}, which
contains two bulk electrodes connected by a constriction.  In order to
understand the dynamical current response, one must formulate the
transport problem beyond the static approach using time-dependent
density functional theory~\cite{Runge,note-closed,Burke} (TDDFT). In
the low frequency limit, the so-called adiabatic local-density
approximation (ALDA) has often been used to treat time-dependent
phenomena in inhomogeneous electronic systems. However, it is
essential to appreciate that the dynamical theory includes an
additional xc field beyond ALDA~\cite{Vignale96} -- {\it a field that
does not vanish in the DC limit}. This extra field, when acting on the
electrons, induces an additional resistance $R^{\rm dyn}$, which is
otherwise absent in the static DFT calculations or TDDFT calculations
within the ALDA~\cite{note-resistance}. Our goal is to find an
analytic expression for this resistance and then estimate its value in
realistic systems. We will show that the dynamical xc field opposes
the purely electrostatic field: one needs a larger electric field to
achieve the same current, implying that the resistance is, as
expected, increased by dynamical xc effects.

We proceed to calculate the xc electric field. This quantity was
determined by Vignale, Ullrich and Conti using time-dependent
current-density functional theory (TDCDFT)~\cite{Vignale97,Ullrich}.
In their notation, this field is
\begin{equation}
E_{\alpha}({\bf r},\omega) = E_{\alpha}^{ALDA}({\bf r},\omega)-\frac{1}{en}
\partial_{\beta} \sigma_{\alpha \beta}({\bf r},\omega) ,
\label{field}
\end{equation}
where $E_{\alpha}^{ALDA}({\bf r},\omega)$ is the ALDA part of the xc
contribution, $\sigma_{\alpha\beta}({\bf r},\omega)$ is the xc stress
tensor, and $e$ is the electron charge. Here $n = n_0({\bf r})$ is the
ground-state number density of an inhomogeneous electron liquid.  We
are interested in the dynamical effects that are related to the second
term on the RHS of Eq.~(\ref{field}), i.e.,
\begin{equation}
E_{\alpha}^{\rm dyn}({\bf r},\omega) \equiv -\frac{1}{e n}
\partial_\beta \sigma_{\alpha\beta}({\bf r},\omega) .
\label{Edyn}
\end{equation}
In the static limit~\cite{Vignale97}, we can transform $E_\alpha^{\rm
dyn}$ into an associated potential by integrating between points $a$
and $b$ inside the electrodes lying on the $z$ axis.  We take the
$z$-direction to be parallel to the current flow.  The end points $a$
and $b$ are in regions where the charge density does not vary
appreciably, i.e., $\partial_z (1/n)|^b_a\sim 0$ (see
Fig.~\ref{model}). This yields
\begin{eqnarray}
\nonumber
V^{\rm dyn}& = & -\int_a^b \lim_{\omega \to 0} \, \mathrm{Re} \, {\bf
E}^{\rm dyn} \cdot d {\bf l} \\*
& = & \int_a^b \frac{1}{e n} \lim_{\omega \to 0} \, \mathrm{Re} \,
\partial_\beta \sigma_{z\beta}({\bf r},\omega) dz .
\label{Vdyn}
\end{eqnarray}
Importantly, we include here only the part of the electric field that
varies in phase with the current (i.e., we take the real part of the
stress tensor).  In general---e.g., at finite frequency---the electric
field has both an in-phase (dissipative) and an out-of-phase
(capacitive) component, where the latter is related to the shear
modulus of the inhomogeneous electron liquid.  Such capacitive
components play a crucial role in the theory of the dielectric
response of insulators. We ignore them here on the basis that they do
not contribute to the resistance~\cite{note-capactive}.

The general xc stress tensor in TDCDFT~\cite{Ullrich} is given by
\begin{eqnarray}
\nonumber
\sigma_{\alpha\beta}({\bf r},\omega) & = & \tilde{\eta}(n,\omega) \left(
\partial_\beta u_\alpha + \partial_\alpha u_\beta - \frac{2}{3}
\nabla \cdot {\bf u}\, \delta_{\alpha\beta}\right)  \\*
&   &  + \tilde{\zeta} \, (n,\omega) {\bf \nabla} \cdot {\bf u} \,
\delta_{\alpha\beta} ,
\label{xcstress}
\end{eqnarray}
where $\tilde{\eta}(n,\omega)$ and $\tilde{\zeta}(n,\omega)$ are the
frequency-dependent viscoelastic coefficients of the electron liquid,
while ${\bf u}={\bf j}/n$ and ${\bf j}$ are the velocity field and the
particle current density, respectively, induced by a small,
time-dependent potential.

The viscoelastic coefficients are given by
\begin{equation}
\tilde{\eta}(n,\omega)=-\frac{n^2}{i \omega} f^h_{xc,T}(\omega)
\end{equation}
and
\begin{equation}
\tilde{\zeta}(n,\omega)=-\frac{n^2}{i \omega} \left\{
f^h_{xc,L}(\omega) - \frac
{3}{4}f^h_{xc,T}(\omega) - \epsilon_{xc}^{\prime \prime} \right\} ,
\end{equation}
where $f^h_{xc,L}(\omega)$ and $f^h_{xc,T}(\omega)$ are, respectively,
the longitudinal and transverse xc kernel of the homogeneous electron
gas evaluated at the local electron density $n=n_0({\bf r})$, while
$\epsilon_{xc}^{\prime \prime}$ is simply
\begin{equation}
\epsilon_{xc}^{\prime \prime} = \left. \frac{d^2 \epsilon_{xc}
(n)}{dn^2} \right|_{n_0({\bf r})} .
\end{equation}

In the representative systems that we examine below, the derivatives
in the transverse directions $x$ and $y$ account for only a small
fraction of the total dynamical xc field and can hence be ignored. We
thus obtain
\begin{equation}
E_z = -\frac{1}{e n} \partial_z \sigma_{zz}.
\end{equation}
We then see that
\begin{equation}
\sigma_{zz}=\frac{4 \eta}{3}\partial_z u_z
\end{equation}
where the viscosity $\eta = \lim_{\omega\to 0} {\rm
Re}\tilde{\eta}(\omega)$ is a function of $n$, and therefore of $z$.
The real part of $\tilde{\zeta}(\omega)$ vanishes in the limit of
$\omega \to 0$.  Under the same assumptions of negligible transverse
variation in current density~\cite{note-j}, we can write
\begin{equation}
u_z = \frac{I}{e n A_c}
\end{equation}
where $I>0$ is the total current (independent of $z$), and $A_c$ is
the cross sectional area~\cite{explain}.  Substituting this into the
equation for the voltage drop and integrating by parts we arrive at
\begin{equation}
V^{\rm dyn}= -\frac{4I}{3 e^2 A_c} \int_a^b \eta \frac{(\partial_z 
n)^2}{n^4}dz~.
\label{voltagedrop}
\end{equation}
Because $\eta$ is positive -- a fact that follows immediately
from the positivity of energy dissipation in the liquid -- we see
that the right hand side of this equation is negative-definite: the
electrostatic voltage is {\it always opposed} by the dynamical xc
effect. We identify the quantity on the right hand side of
Eq.~(\ref{voltagedrop}) with $- R^{\rm dyn} I$, where $R^{\rm dyn}$ is
the additional resistance due to dynamical effects.
According to TDCDFT, the current that flows in the structure in
response to an electric potential difference $V$ is given by
\begin{equation}
\label{current}
I = G_{s}(V+V^{\rm dyn}) =  G_{s}(V-R^{\rm dyn}I)
\end{equation}
where $G_s$ is the conductance calculated in the absence of dynamical
xc effects (e.g., by means of the techniques of
Ref.~\onlinecite{DiVentra01}). Solving Eq. (\ref{current}) leads
immediately to the following expression for the total resistance
$R=V/I$:
\begin{equation}
R =
\frac{1}{G_s} + R^{\rm dyn}~,
\end{equation}
where
\begin{equation}
R^{\rm dyn} = \frac{4}{3 e^2 A_c} \int_a^b \eta\frac{( \partial_z n
)^2 }{n^4} dz.
\label{Rdyn}
\end{equation}
The dynamical xc term thus {\it increases} the resistance.

This is the central result of our paper. It shows that the dynamical
effects (beyond ALDA) on the resistance depend nonlinearly on the
variation of the charge density when the latter changes continously
from one electrode to the other across the junction. In a
nanojunction, this correction is non-zero only at the
junction-electrode interface where the electron density varies
most. Knowing the charge density one can then estimate this
resistance.

\begin{figure}
\includegraphics*[width=8.0cm]{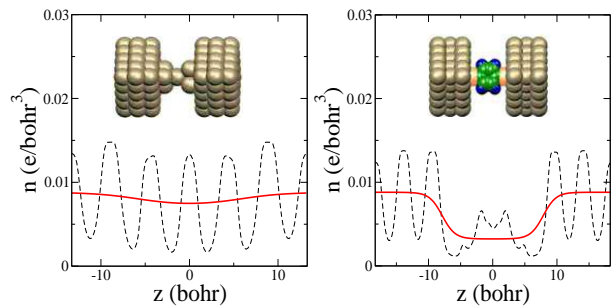}
\caption{Planar (dashed line) and macroscopic (solid line) averages of
the charge density for a gold point contact (left) and a molecular
junction (right). }
\label{DFT_den}
\end{figure}
Let us thus consider two specific examples that have attracted much attention,
namely the gold point contact and the molecular junction formed by a
benzene-dithiolate (BDT) molecule between two bulk electrodes
(see insets in Fig.~\ref{DFT_den}) and estimate the error made by the 
(A)LDA calculation in
determining the resistance. In order to make a connection
between the microscopic features of these junctions and the density in
Eq.~(\ref{Rdyn}), we model the charge density $n=n_0({\bf r})$ as
a product of smoothed step functions in every direction, i.e.,
\begin{eqnarray}
n_0({\bf r}) & = & n_e \Xi(z,d,\gamma_e) \nonumber\\ & & + n_c
               \Omega(x,h,\gamma_c)
               \Omega(y,w,\gamma_c)\Omega(z,d,\gamma_e) .
\label{density}
\end{eqnarray}

The smoothed step function is given by
\begin{equation}
\Theta(z,l,\gamma) = \frac{1}{e^{(z+l/2)/\gamma}+1} ,
\end{equation}
where $l$ is the full-width at half-maximum and $\gamma$ is the decay
length. Here, $\Omega(z,l,\gamma) = \Theta(z,-l,\gamma) -
\Theta(z,l,\gamma)$ and $\Xi(z,l,\gamma) = 1-\Omega(z,l,\gamma)$;
$\Omega(z,l,\gamma)$ represents the density distribution of the
junction $n_c$, which smoothly connects to the constant bulk density
$n_e$ of the two electrodes separated by a distance $d$. Finally, $h$
and $w$ represent the lateral dimensions of the junction.

The electron densities are obtained from self-consistent static DFT
calculations with the xc functional treated within the
LDA~\cite{Abinit}. The (111) gold surface orientation is chosen for
both the point contact and the molecular junction (see schematics in
Fig.~\ref{DFT_den}).

\begin{figure}
\includegraphics*[width=5.5cm]{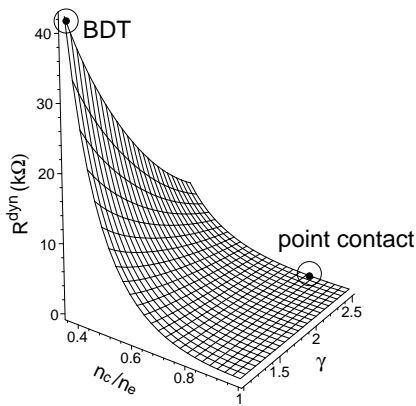}
\caption{The resistance due to dynamical effects as
calculated from Eq.~\ref{Rdyn} with the charge density determined from
DFT-LDA calculations as a function of the main parameters discussed in
the text. The resistance of a gold point contact and a BDT molecular
junction are indicated by dots. }
\label{resistance}
\end{figure}

In Fig~\ref{DFT_den} we plot the planar and macroscopic averages of
the self-consistent electron densities for both systems as a function
of distance from the center of the junction along the
$z$-direction. The macroscopic average is then fitted to the simple
charge model in Eq.~(\ref{density}).  The fitted density is then
substituted in Eq.~(\ref{Rdyn}) to find the correction to the
resistance. The estimated value of $R^{\rm dyn}$ \cite{note-eta} for
the point contact is $\sim$0.2 K$\Omega$ (the static DFT resistance 
is about 12 K$\Omega$~\cite{Lager}), while for the BDT molecule
is $\sim$40K$\Omega$ (to be compared with the linear response static DFT resistance of 
about 360 K$\Omega$~\cite{DiVentra01}). As expected, $R^{\rm dyn}$ for BDT is larger
than that for the point contact due to the larger variation of the
average density between the bulk electrodes and the molecule. In
Fig.~\ref{resistance} we plot the resistance in Eq.~(\ref{Rdyn}) as a
function of the ratio $n_e/n_c$ and the decay constant $\gamma$, where
we fix $n_e$ to the value of bulk gold ($r_s \approx 3$). The
resistances of the two specific examples are indicated by dots in the
figure. It is clear that the dynamical contributions to
the resistance can become substantial when the gradient of the density
at the electrode-junction interface becomes large.  These corrections
are thus more pronounced in organic/metal interfaces than in
nanojunctions formed by purely metallic systems.

In summary, we have shown that dynamical effects in the xc
potential contribute an additional resistance on top of the
static DFT-LDA one. The magnitude of the additional resistance, within
linear response and the zero-frequency limit, depends on the gradient
of the charge density across the junction. This additional resistance
is thus larger in molecular junctions than in quantum point contacts.

We acknowledge financial support from an NSF Graduate Fellowship (MZ),
  NSF Grant No.~DMR-0313681 (GV) and NSF Grant No. DMR-0133075 (MD).


\begin{references}
\bibitem{note-resistance} We implicitly assume that we can identify a
physical initial state such that, in the zero-frequency limit, the
conductance obtained from a TDDFT calculation within the ALDA for the
xc kernel equals the conductance obtained using static DFT within
LDA. The assumption requires non-resonant scattering and is thus
satisfied by quantum point contacts and molecular junctions in linear
response when the Fermi level falls between the highest occupied and
the lowest unoccupied molecular orbitals~\cite{DiVentra00}.
\bibitem{note-capactive} In the current-density functional theory, the
capacitive effect is described by a static shear modulus. Within the
local density approximation for a metallic system, this quantity is
expected to vanish in the frequency regime relevant to DC conduction.
\bibitem{explain} Note that we have assumed the right electrode is
positively biased (see Fig.~\ref{model}) so that electrons flow from
left to right. The current is thus defined as flowing in the opposite
direction.
\bibitem{note-eta} We have used the viscosity $\eta =
\frac{\hbar}{120} \left[\frac{k_F}{\pi a_0}\right]^{3/2}$ where $k_F$
and $a_0$ are the Fermi wave-vector of the bulk electrodes and Bohr
radius, respectively.
\end{references}
\end{document}